\documentclass[12pt]{article}

\usepackage{authblk}
\usepackage{scicite}
\usepackage{times}
\usepackage{amsmath}
\usepackage{amssymb}
\usepackage[dvipsnames,svgnames,table]{xcolor}
\usepackage{graphicx}
\usepackage{physics}
\usepackage[hidelinks]{hyperref}
\usepackage[left,mathlines,running]{lineno}
\definecolor{bristolred}{RGB}{191,47,56}
\definecolor{darkred}{RGB}{192,0,0}
\definecolor{darkgreen}{RGB}{64,128,32}
\definecolor{darkblue}{RGB}{0,0,128}

\def\btext#1{{\color{darkblue}#1}}

\def\epsilon{\varepsilon}

\newcommand{\eq}[1]{\begin{equation}#1\end{equation}}

\newif\iflectureversion
\lectureversionfalse

\renewcommand{\vec}[1] {\boldsymbol{#1}}
\newcommand{\mat}[1]{{\bf #1}}

\newcommand{\al}[1]{\begin{align}#1\end{align}}

\newcommand{\eqa}[1]{\begin{subequations}
\begin{align}#1\end{align}\end{subequations}}

\newenvironment{sciabstract}{%
\begin{quote} \bf}
{\end{quote}}

\usepackage{hyperref}
\usepackage{soul}
\newcommand{\hrefc}[3][blue]{\href{#2}{\color{#1}{#3}}}
\definecolor{myLink}{HTML}{08519c}

\usepackage{sectsty}
\definecolor{mySect}{HTML}{333333}
\sectionfont{\large\sc\color{mySect}}
\subsectionfont{\normalsize\sc\color{mySect}}

\usepackage[backend=biber,style=science]{biblatex}
\usepackage{fancyhdr}
\usepackage{lastpage}
\def\scititle{
	Functional Motifs in Foodwebs and Networks
}

\title{Functional Motifs in Foodwebs and Networks} 

\author[1,2,3,$\star$]{\hrefc[myLink]{https://orcid.org/0009-0006-4733-6769}{Melanie Habermann}}
\author[4]{\hrefc[myLink]{https://orcid.org/0000-0001-9138-3593}{Ashkaan K. Fahimipour}}
\author[5,6]{\hrefc[myLink]{https://orcid.org/0000-0002-6597-3511}{Justin D. Yeakel}}
\author[1,2,3]{\hrefc[myLink]{https://orcid.org/0000-0002-1356-6690}{Thilo Gross}}

\affil[1]{Helmholtz Institute for Functional Marine Biodiversity (HIFMB), Im Technologiepark 5, 26129 Oldenburg, GER}
\affil[2]{Alfred-Wegener Institute (AWI), 27075 Bremerhaven, GER}
\affil[3]{Carl-von-Ossietzky Universit\"{a}t, ICBM, 26129 Oldenburg, GER}
\affil[4]{Florida Atlantic University, 777 Glades Rd., Boca Raton, FL, USA}
\affil[5]{University of California Merced, 5200 North Lake Rd., Merced, CA, USA}
\affil[6]{The Santa Fe Institute, 1399 Hyde Park Rd., Santa Fe, NM, USA}

\date{}

\begin{document} 
\sloppy
\baselineskip24pt
\maketitle 
\thispagestyle{fancy}

\begin{sciabstract}
When studying a complex system it is often useful to think of the system as a network of interacting units. One can then ask if some properties of the entire network are already explained by a small part of the network--a network motif. A famous example of an ecological motif is competitive exclusion in foodwebs, where the presence of two species competing for a shared resource precludes the existence of a stable equilibrium for the whole system. However, other examples of motifs with such direct impacts on stability are not known. Here we discuss why small motifs that allow conclusions on systemic stability are rare. More importantly, we show that another dynamical property, reactivity, is naturally rooted in motifs. Computing the reactivity of motifs can reveal which parts of a network are prone to respond violently to perturbations. This highlights motif reactivity as a useful property to measure in real-world systems to understand likely modes of systemic failure in foodwebs, epidemics, supply chains, and other applications.  
\end{sciabstract}

\noindent
In its modern form the competitive exclusion principle states that the number of species coexisting in an ecological system cannot exceed the number of realized niches \cite{gause2019struggle, hardin1960competitive}. Phrased thus, the principle is tautological \cite{levin1970community}, as a niche is defined as circumstances that allow a species to persist. This tautological nature makes the competitive exclusion principle an inviolable law, whose apparent violation in nature has guided the search for overlooked niches \cite{macarthur1958population, gross2009invisible, mcpeek2012intraspecific}, leading to major discoveries \cite{macarthur1958population,armstrong1980competitive, chesson1988interactions, siepielski2010evidence}.

The success of the competitive exclusion principle is rooted in its applicability to small, isolated parts of an ecological network, avoiding the complexity that exists in the wider system. The simplest illustrative case is the \emph{exploitative competition motif}, where two unregulated consumers specialize on one resource \cite{tilman1982resource}. Hence, the presence of this motif in a foodweb implies that some (perhaps so-far undetected) internal regulation of the consumers must be present \cite{abrams2001effect, gross2009invisible}. Importantly this result is independent of the structure of the rest of the network, which makes the principle a valuable tool for ecology. It's thus reasonable to ask if there are other network motifs with similar predictive power, which could lead to new principles as powerful as competitive exclusion. Although stable motifs have a tendency to stabilize foodwebs \cite{borrelli2015subgraphs, cirtwill2022stable}, even extensive numerical search has not identified other motifs that have as clear-cut implications as the exploitative competition motif \cite{stouffer2010understanding}.

Here we use the term \emph{functional motif} to indicate a network motif that by its mere presence has implications that cannot be negated by the rest of the network. We explain why exploitative competition is a functional motif with respect to the stability of ecological states and also show why other such motifs are unlikely to exist. We then discuss other mechanisms by which motifs can become functional in any type of network. Of particular interest is reactivity, which provides a different notion of ecological stability \cite{neubert1997alternatives}; we show that every network motif is a functional motif with respect to reactivity. This highlights reactivity of parts of a system as an important property that should be measured in real-world networks.   


\section*{Functional Stability Motifs}
Consider a generic many-variable dynamical system whose dynamics are described by a system of differential equations. We can picture such a system as a network, where nodes correspond to variables and weighted directed links indicate the interaction between pairs of variables.  

A common notion of stability, locally asymptotic stability, can be computed by studying so-called eigensolutions of the dynamics in the proximity of a dynamical behavior \cite{guckenheimer1990nonlinear}. In the simplest case the behavior under consideration is stationary in a steady state. The dynamics in the proximity are then captured by the Jacobian matrix, $\bf J$, and the eigensolutions are given by the eigenvectors and eigenvalues of this matrix, found by solving the eigenvalue equation
\eq{{\bf J}\vec{v}=\lambda \vec{v},}
where $\vec{v}$ is an eigenvector and $\lambda$ the corresponding eigenvalue. Specifically, a stationary state is stable if all solutions for $\lambda$ are negative. Similar conditions also exist for non-stationary states but are in practice harder to compute and hence less frequently evaluated.  

Here we address the conditions under which analyzing a small network motif is sufficient to make statements about system stability. Since the presence of a single non-negative Jacobian eigenvalue makes a system unstable, a motif is a functional stability motif if its presence in a system implies that at least one eigenvalue will have a non-negative real part.

\begin{figure}[ht!]
  \centering
  \includegraphics[width=0.9\textwidth]{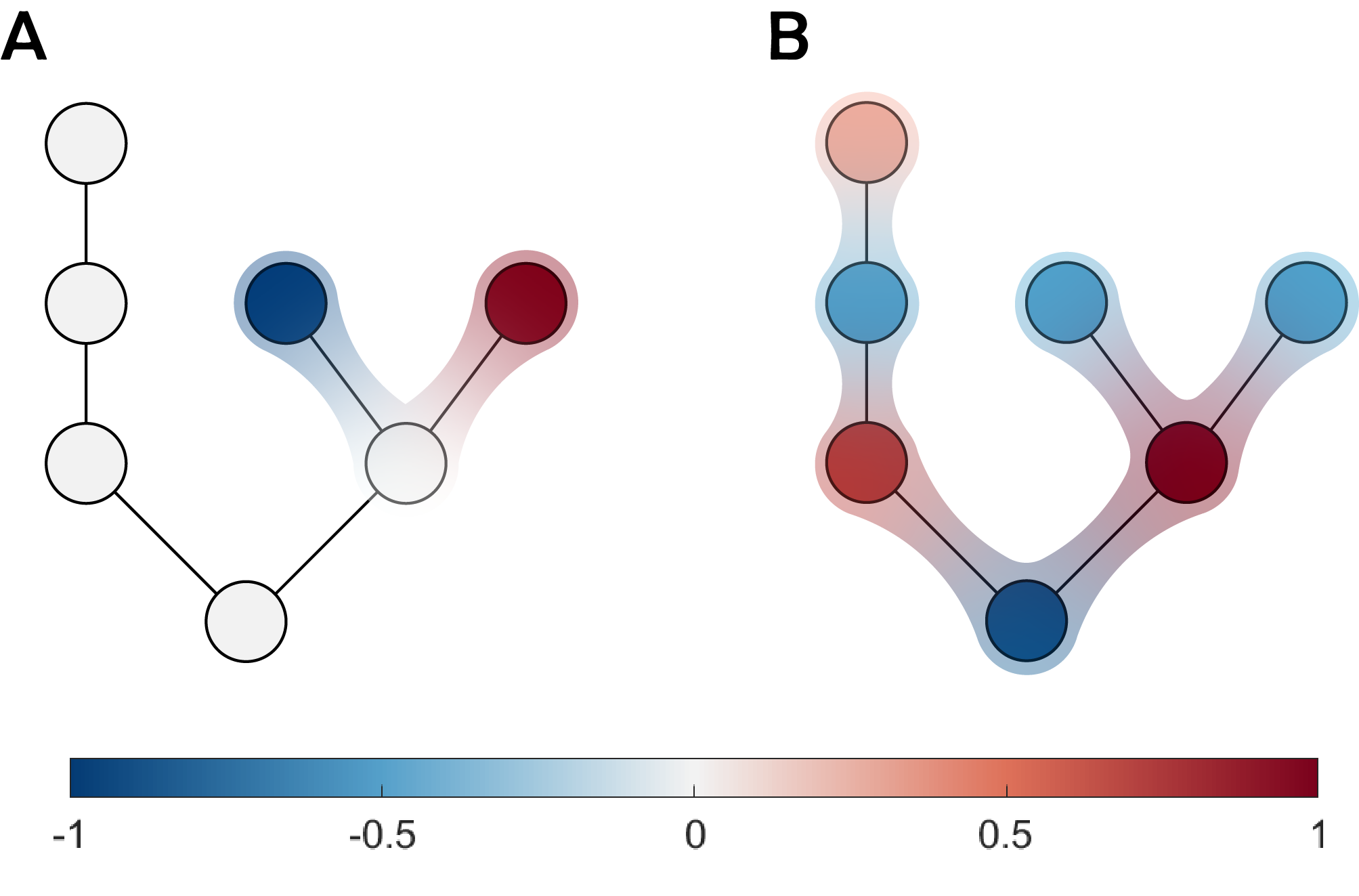}
  \caption{\textbf{Comparison between localized and delocalized eigenvectors of a Jacobian matrix.} The example shows two different eigenvectors of the Jacobian matrix from the same system as heatmaps. Nodes corresponding to zero elements (white) in the eigenvector have no impact on the associated eigenvalue. The exploitative competition motif localizes an eigenvector (panel A), such that the corresponding eigenvalue becomes independent of nodes outside the motif. By contrast typical eigenvectors (panel B) which also exist in the same system are delocalized and hence depend on all of the nodes. This illustrates that eigenvector localization, if it occurs in a system, can result in the formation of functional motifs.
}
  \label{fig1:localev}
\end{figure}

One can imagine two different ways in which the presence of a specific motif could imply dynamical instability: First, a given motif could imply that there must always be one eigenvalue that has a specific value. Second, a given motif could imply that there must be a positive eigenvalue, without specifying this eigenvalue precisely. In the following we refer to these cases respectively as functional motifs of the first and second kind.  

We start by considering motifs of the first kind, where the presence of a motif directly causes the presence of a specific eigenvalue. Generally, the eigenvalues of a large matrix don't originate from a small motif but emerge from the matrix as a whole (Fig.~\ref{fig1:localev}). To explore this more deeply, we can build on results from matrix perturbation theory \cite{stewart1990matrix}, which show that the set of nodes that `cause' an eigenvalue is revealed by the corresponding eigenvector. For a positive eigenvalue, the corresponding eigenvector plays a two-fold role. First, the eigenvector is tangential to the unstable manifold of the state and thus indicates a direction in which a system can escape from the state \cite{guckenheimer1990nonlinear}. Second, and more importantly, the eigenvector indicates to which nodes the eigenvalue is sensitive \cite{stewart1990matrix}. 

In a system of $N$ variables, Jacobian eigenvectors contain $N$ elements, each corresponding to one of the variables. If a normalized eigenvector has a large element, then the corresponding variable has a strong impact on the associated eigenvalue. By contrast a small element means that the corresponding variable has only a weak impact and a zero element means that an eigenvalue is (at least locally) independent of the variable. We can thus say that an eigenvector has a functional stability motif of the first kind if there is a positive eigenvalue for which the corresponding eigenvector is zero in all nodes outside the motif.

\section*{Exploitative competition example}
Common derivations of the competitive exclusion principle typically start by considering a system of two populations of specialist competitors of the form 
\eqa{
\dot{X} &= G_{\rm x} (R)X  - m_{\rm x} X \\
\dot{Y} &= G_{\rm y} (R)Y  - m_{\rm y} Y  
\label{eqCE}
}
where $X$ and $Y$ are abundances or biomasses of the two populations, $m$ are the respective mortality rates, $G$ are the growth rates, and the dot over a variable denotes the change in time. The variable $R$ is the resource that both of the competing consumer populations are exploiting and which possibly interacts with many other species that do not have a direct impact on the competing consumers. The dynamics of $R$ and any other variables in the system are not essential to the argument that is made and are hence omitted here.

One can show that the coexistence of the species is only feasible if one of the parameters (\emph{e.g.}, $m_{\rm x}$) is chosen exactly right \cite{tilman1982resource, abrams1983theory}, which is thought to be implausible in nature. However, this reasoning has been criticized as it applies a genericity argument to a system where the degeneracy is rooted in the modeling assumptions \cite{gross2009invisible} (see supporting information,  SI).

Let's instead consider the stability of the steady state if one exists. For any such steady state the eigenvector equation can be solved as: 
\eq{
\underbrace{\left[\begin{array}{c c c c c c} 
  \btext{0} & \btext{0} & a_1 & 0 & \cdots \\
  \btext{0} & \btext{0} & a_2 & 0 & \cdots \\
  b_1 & b_2 & c_{11} & c_{12} & \ldots \\
  0 & 0 & c_{21} & c_{22} & \ldots \\
  \vdots & \vdots &  \vdots & \vdots & \ddots 
\end{array} 
\right]}_{\bf J} 
\underbrace{
\left(\begin{array}{c} 
b_2 \\ -b_1 \\ 0 \\ 0 \\ \vdots 
\end{array} 
\right)
}_{\vec{v}}
=
\underbrace{
\left(\begin{array}{c} 
0 \\ 0 \\ 0 \\ 0 \\ \vdots 
\end{array} 
\right)
}_{\lambda \vec{v}},
}
where the first columns correspond to the variables $X$, $Y$, the third corresponds to the resource $R$ and all others describe other variables in the system. The Jacobian matrix has a block of zeros (blue) in the top-left as the two populations X and Y do not interact directly and self-interactions vanish as a consequence of the linearity of gain and loss rates with respect to the consumer populations \cite{gross2006gm}. Moreover, the two competitors only interact with the rest of the network via their resource $R$ (third row). The only non-zero elements that we find in the first two columns of the matrix are hence $b_1$ and $b_2$, which describe the consumers' impact on the resource and whose values will depend on the specific system under consideration. However, a vector of the form $\vec{v}$ shown in the equation will always be an eigenvector of the matrix with a corresponding eigenvalue $\lambda=0$.    

The example shows that the exploitative competition motif implies the presence of a zero eigenvalue, regardless of the interactions in the rest of the system (represented by $c_{ij}$). Hence, any system involving competition between uncontrolled consumers described by Eq.~\ref{eqCE} always has a zero eigenvalue, which precludes asymptotic stability (see SI for details). We can therefore say that the two species whose dynamics are described by Eq.~\ref{eqCE} form a functional stability motif, as their joint presence in the absence of internal regulation inherently disrupts stability. 

Note that the feasibility of stationary coexistence is tied to the presence of this zero eigenvalue through the implicit function theorem of calculus, which allows us to compute how stationary states change as parameter values shift. If the Jacobian has a zero eigenvalue, the solutions diverge, signaling an abrupt change in the solution and typically annihilation of the equilibrium.

\section*{Functional stability motifs are rare}
The exploitative competition example illustrates that an eigenvalue is determined by a small motif if the eigenvector elements that correspond to nodes outside the motif are zero. Such eigenvectors that are zero except in a small motif are called \emph{localized eigenvectors}. We can now state that a functional stability motif of the first kind requires the presence of a localized eigenvector. 

In connected networks \emph{exact localization}, where eigenvector elements outside a motif become exactly zero, appears only in response to the presence of certain symmetries in the network \cite{macarthur2009spectral}. In unweighted networks, these symmetries and hence localized eigenvalues are common \cite{nyberg2015mesoscopic}. However, in network representations of Jacobians, which are weighted networks, link weights must also satisfy certain symmetry conditions to allow for eigenvector localization. Specifically a vector is a localized eigenvector for a specific motif if: [\emph{i}] it is an eigenvector for the motif in isolation, and [\emph{ii}] the eigenvector elements are zero on those nodes in the motif to which outside nodes are allowed to attach. 

For example, consider the modified exploitative competition system 
\eqa{
\dot{X} &= F(R)X - m X^{1+p}\\ 
\dot{Y} &= F(R)Y - m Y^{1+p}
}
which has a stationary state with $X^*=Y^*>0$. The Jacobian matrix for the motif in isolation, including $R$, has the structure 
\eq{
J_{[1:3]} = \left[ \begin{array}{ccc} d & 0 & a \\
0 & d & a \\  b & b & c \end{array}  \right],
}
where $a,b,c,d$ depend on specifics of the system. For this matrix the vector $(1,-1,0)'$ is an eigenvector and the element corresponding to $R$ is zero. Hence, this motif is a functional motif of the first kind as long as other parts of the network only attach to the resource node. 

The eigenvalue corresponding to the localized vector is $\lambda = d = -mp (X^\star)^p$. Hence, it's presence in the network will destabilize the entire system if $p \leq 0$, but not if $p>0$, which is a known result \cite{gross2009invisible}.

In this example the localization appears because we assumed the two competitors behave identically, preserving complete symmetry between the species\cite{AufderheideThesis}. While it's possible to find nonsymmetric examples where fortuitous cancellations happen to meet the conditions, these are likewise special cases. When searching for a vector that meets the two conditions, the condition (i) alone reduces the choice down to a narrow set of candidates, as the motif in isolation only has a finite set of eigenvectors. The probability that any of these eigenvectors is able to meet condition (ii) exactly is of measure zero, except if a symmetry exists. The exploitative competition motif with linear rates escapes the logic above because the motif block contains only zeroes and every vector is an eigenvector of a zero matrix, which makes it easy to meet the second localization condition.  

\section*{Routes to functional stability motifs}
There are some additional ways in which functional stability motifs of the first kind can arise. While the implications for ecology are mostly well-known, we find it valuable to provide a brief but comprehensive overview. We have seen that the original exploitative competition motif is a functional motif because it assumes {\bf linear rates} and {\bf absence of internal interactions} within the motif. This will create a functional motif whenever we have the sufficient degrees of freedom left to satisfy the condition on adjacent nodes, \emph{i.e.}, whenever the number of nodes within the motif is greater then the number of adjacent nodes, which is equivalent to the general formulation of the competitive exclusion principle. 

Our second example was a functional motif due to the assumption of {\bf identical dynamics}. A motif is a functional motif if there is a non-trivial symmetry in the Jacobian matrix (this is the case whenever the order of variables can be changed while leaving the matrix unchanged, see SI). In practice this will require the assumption that there are sets of variables that behave identically. In ecology this assumption could be warranted as an approximation as there are many examples of species that appear to be functionally redundant 

A trivial way to achieve eigenvector localization, which we only mention for completeness, is {\bf disconnection} or {\bf uni-directional connection}. If a part of a system interacts with the rest of the system not at all or only uni-directionally then also eigenvectors will remain localized in their respective parts. Such an isolation of a part of the system is rare and hence this scenario will be irrelevant in most applications.

If the conditions for exact localization of eigenvectors are not met, we can sometimes still have {\bf Anderson localization} \cite{anderson1958absence}. In this case, eigenvectors from the motif spill into the rest of the network, but the eigenvector elements decay exponentially with increasing distance from the motif, leading to an approximate localization which makes an eigenvalue approximately independent of the nodes outside the motif. 

We expect Anderson localization when the conditions for exact localization are almost met, and the wider network cannot sustain the respective eigenvalue. For example Anderson-localized eigenvectors arise if a tightly knit community within a motif is only weakly interacting with a wider network that is sparser or operating on different time scales. Anderson localization may thus provide a post-factum justification for modeling, say, an aquatic foodweb within a lake separately from the surrounding terrestrial system \cite{fahimipour2014dynamics}, or modeling a mammalian foodweb without considering invertebrates and microbes \cite{yeakel2014collapse}. In these cases we would expect eigenvectors to become Anderson-localized, making the system approximately insensitive to the wider world. Hence, for example, the lake foodweb can become a functional motif in the broader ecosystem.

In summary, the discussion in this section shows that there are a number of scenarios that lead to functional motifs of the first kind. However, these scenarios are relatively rare and have consequences that are already well known, at least within the context of community ecology. Let us therefore now turn to functional motifs of the second kind. While these don't guarantee the presence of a specific eigenvalue, they guarantee that one eigenvalue will have at least a certain value.

In general dynamical systems, the presence of a motif doesn't constrain the Jacobian eigenvalues of the system as a whole. Therefore, functional stability motifs of the second type only exist under specific circumstances. However, these motifs become ubiquitous in the sub-class of dynamical systems that have symmetric Jacobian matrices, meaning $J_{ij}=J_{ji}$ for all $i,j$. Such systems arise for example in the modeling of coupled oscillators, meta-populations, and epidemics on networks.

For a symmetric matrix, application of the eigenvalue interlacing theorem \cite{hwang2004cauchy} proves that the largest eigenvalue of the entire system must be at least as large as the largest eigenvalue of any subgraph that we find in the system. Hence, in systems with symmetric Jacobians any subgraph can act as a functional stability motif. 

\section*{Functional Reactivity Motifs}
\begin{figure}[ht!]
  \centering
  \includegraphics[width=0.9\textwidth]{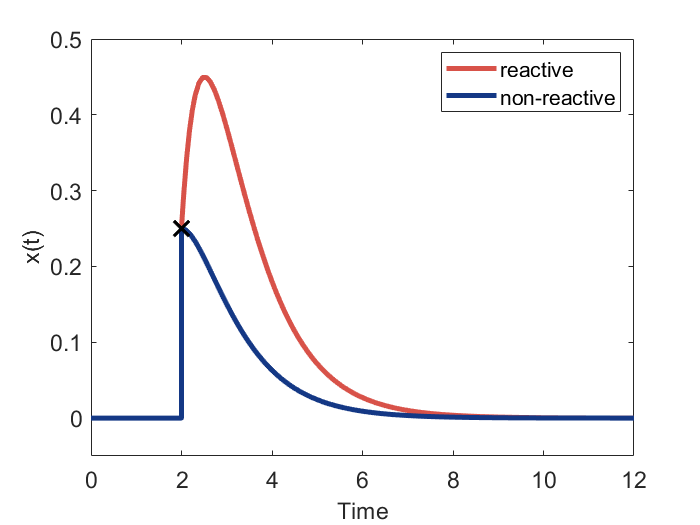}
  \caption{\textbf{Behavior of reactive and non-reactive stable systems.} A system in stable equilibrium responds to a sufficiently small perturbation (cross) by eventually returning to its equilibrium state (here, 0). In non-reactive systems this return is uniform (blue). Conversely, in a reactive system a transient amplification of the perturbation occurs (red).}
  \label{fig2:reactivity}
\end{figure}

Besides local asymptotic stability, many other notions of stability have been proposed in ecology and beyond. Here we consider in particular the notion of reactivity which has attracted considerable interest in ecology \cite{neubert1997alternatives, snyder2010makes, hastings2021effects, yang2023reactivity}. In contrast to asymptotic stability, which focuses on the long-term response to perturbations, reactivity describes the system's initial reaction (Fig.~\ref{fig2:reactivity}). Specifically, a system is said to be reactive if it responds to a perturbation by launching into an excursion that takes it (at least transiently) farther away from its previous state. While every unstable system is reactive, stable systems can be reactive or non-reactive.     

It has been shown mathematically that a system is reactive if the symmetric part of the Jacobian, ${\bf S}=({\bf J}+{\bf J}')/2$, has a positive eigenvalue \cite{neubert1997alternatives}. The largest eigenvalue of the symmetric part of the Jacobian is then said to be the system's \emph{reactivity}, which is an indicator of the factor by which the system amplifies perturbations.  

Because reactivity is computed as an eigenvalue of a symmetric matrix $\bf S$, the interlacing theorem implies that the largest eigenvalue of $\bf S$ must be equal or greater than the largest eigenvalue of any motif found in $\bf S$. In other words: observing a certain amount of reactivity in a motif already provides a lower bound for the reactivity for the entire system, and thus every subgraph of a network is a functional reactivity motif. 

To explore how much of a system's reactivity is explained by small motifs, we studied reactivity in model foodwebs.  
For this purpose we generated plausible foodweb topologies with 15 species using the ecological niche model \cite{williams2000simple}. We then converted the foodweb topologies into a dynamical model using the generalized modeling approach \cite{gross2006gm} (see SI). The result is a set of plausible Jacobian matrices for steady states in foodwebs, depending on a set of parameters. The parameters are then drawn randomly from biologically plausible parameter ranges. We rejected Jacobians corresponding to unstable states, as these would be poor models of states observed in nature. Additionally, we excluded non-reactive stable states, as we are interested in systems where reactivity would be a relevant concern. 

Following the procedure described above, we created two sets containing $10^4$ Jacobians for stable but reactive systems. One set contained random topologies with randomly chosen parameters (set 1) whereas the second set contained Jacobians with randomly chosen parameters for a single fixed network topology (set 2). 

The analyses focused on a set of well-known foodweb motifs including the tri-trophic chain (TC), apparent competition (AC), exploitative competition (EC), and omnivory (O). For both sets of foodwebs, we computed system reactivity as the leading eigenvalue of the symmetric part of the Jacobian. We also identified all occurrences of these motifs and computed their inherent reactivity as the leading eigenvalue of the corresponding blocks in $\bf S$. 

\begin{figure}[ht!]
  \centering
  \includegraphics[width=0.9\textwidth]{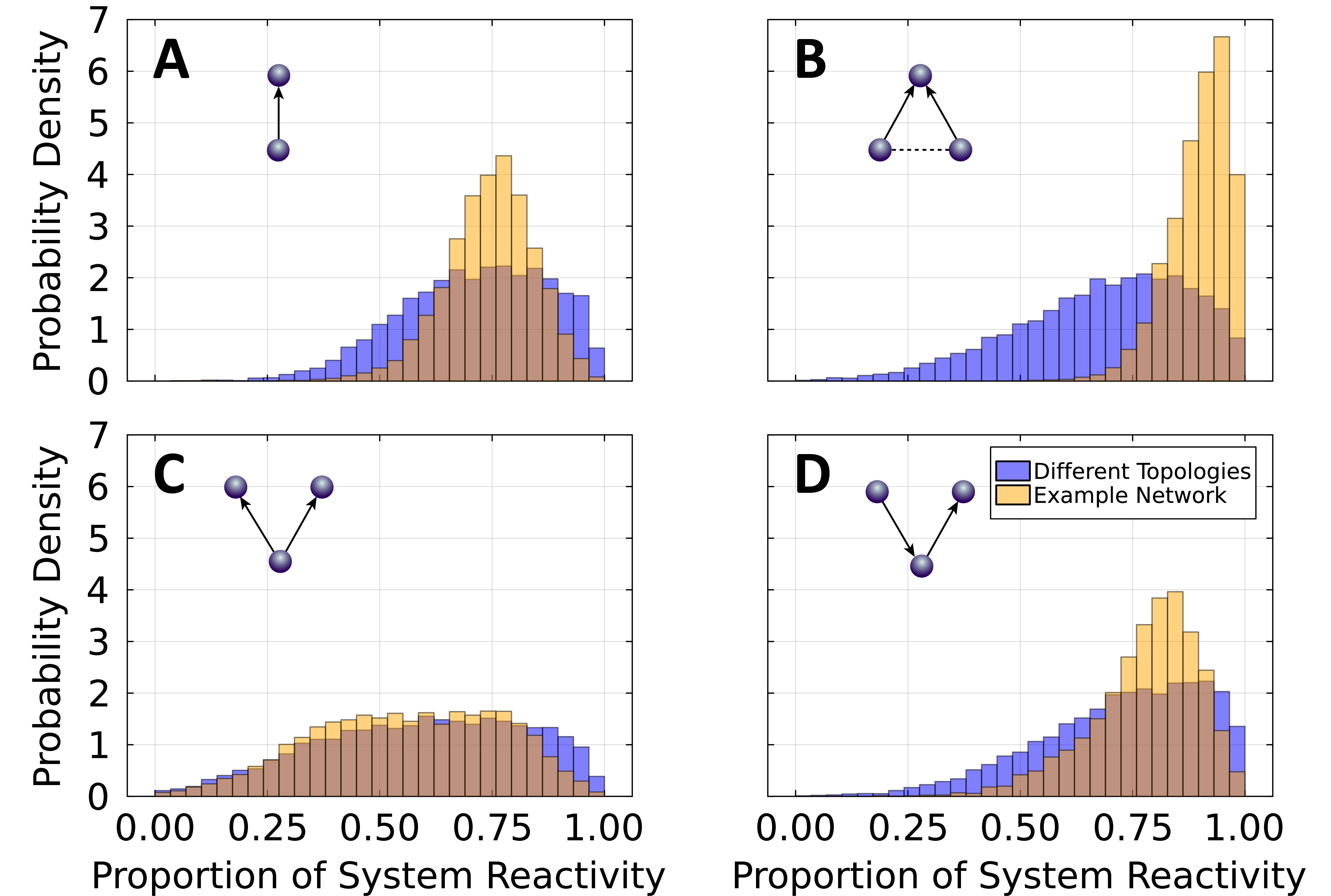}
  \caption{\textbf{Contribution of individual motifs to system reactivity.} The histograms show the proportion of total system reactivity contained in a single motif across $10^4$ simulations. The respective motives predator-prey motif (PP, panel A), apparent competition motif (AC, B),  exploitative competition motif (EC, C), and tri-trophic chain (TC, D) are indicated in the top left corners. For a 15-node example network with different parameterizations (orange), the most reactive apparent competition motif already contains a significant fraction of the entire network's reactivity. In an ensemble of networks with different topologies (blue), PP, AC, and TC contribute very strongly to reactivity compared to the EC motif. These results illustrate that it is valuable to evaluate the reactivity of parts of a larger network. The same comparison for additional motifs can be found in the SI Appendix in Fig. $S1$.}
  \label{fig4:comp_reactprop}
\end{figure}

To assess whether small motifs can explain a significant proportion of system-wide reactivity, we examined the most reactive instance of each motif and compared its reactivity to that of the entire system (Fig.~3). In the most extreme case we find that the most reactive predator-prey link alone accounts for $\geq 99.9 \%$ of the total network reactivity in set 1 and $\geq 99.3 \%$ in set 2. This result is surprising given that reactivity can only increase with motif size. Furthermore, in all but $10.7\%$ of cases in set 1 and only $1.8\%$ in set 2, the most reactive link contributes at least half of the system's reactivity.

Expanding the analysis to three-node motifs reveals a wide range of relative contributions to system reactivity. In some cases, the most reactive motif accounts for $\geq 99.9 \%$ of the total reactivity for both sets. In the set of random topologies (set 1), the TC often accounts for a significant proportion of system-wide reactivity. In $\geq 86.7 \%$ of all cases, the TC motif alone can explain more than $50 \%$ of system reactivity and in $\geq 49.7 \%$ of all cases for at least $75 \%$. We could find similar good results with the AC and O motifs, where $\geq 82.7 \%$ and $\geq 83.0 \%$ of all computed communities could account for more than half of total reactivity.

EC is a comparatively weaker predictor of system reactivity. Unregulated exploitative competition would result in instability and is not considered here, while strongly regulated exploitative competition motifs contribute less to reactivity. Interestingly, even though EC showed itself to be a weaker predictor when taking all communities into account, there were still extreme cases where it turned out as the most reactive motif and accounted for $99.9 \%$ of system reactivity. 

In the specific example topology (set 2), AC emerges as the most influential three-node motif. In more than half of all cases ($55.7 \%$), the most reactive AC motif accounted for $\geq 90 \%$ of the system's total reactivity. 

The same analysis for motifs with four and five nodes can be found in Fig. $S1$ in the SI appendix. Although one could assume that they are better predictors of system reactivity based on the increasing motif size and with it theoretically higher reactivity, this is not necessarily the case. Larger motifs are less likely to find and don't have to include the most reactive smaller motif. If this is the case, it is comprehensible that one of the smaller motifs can be more reactive and with this a better approximation of system reactivity.

This suggests that identifying small reactive motifs may be especially useful in real-world systems where reactivity presents a significant concern \emph{e.g.}, \cite{yang2023reactivity}.

\begin{figure}[t!]
  \centering
  \includegraphics[width=0.9\textwidth]{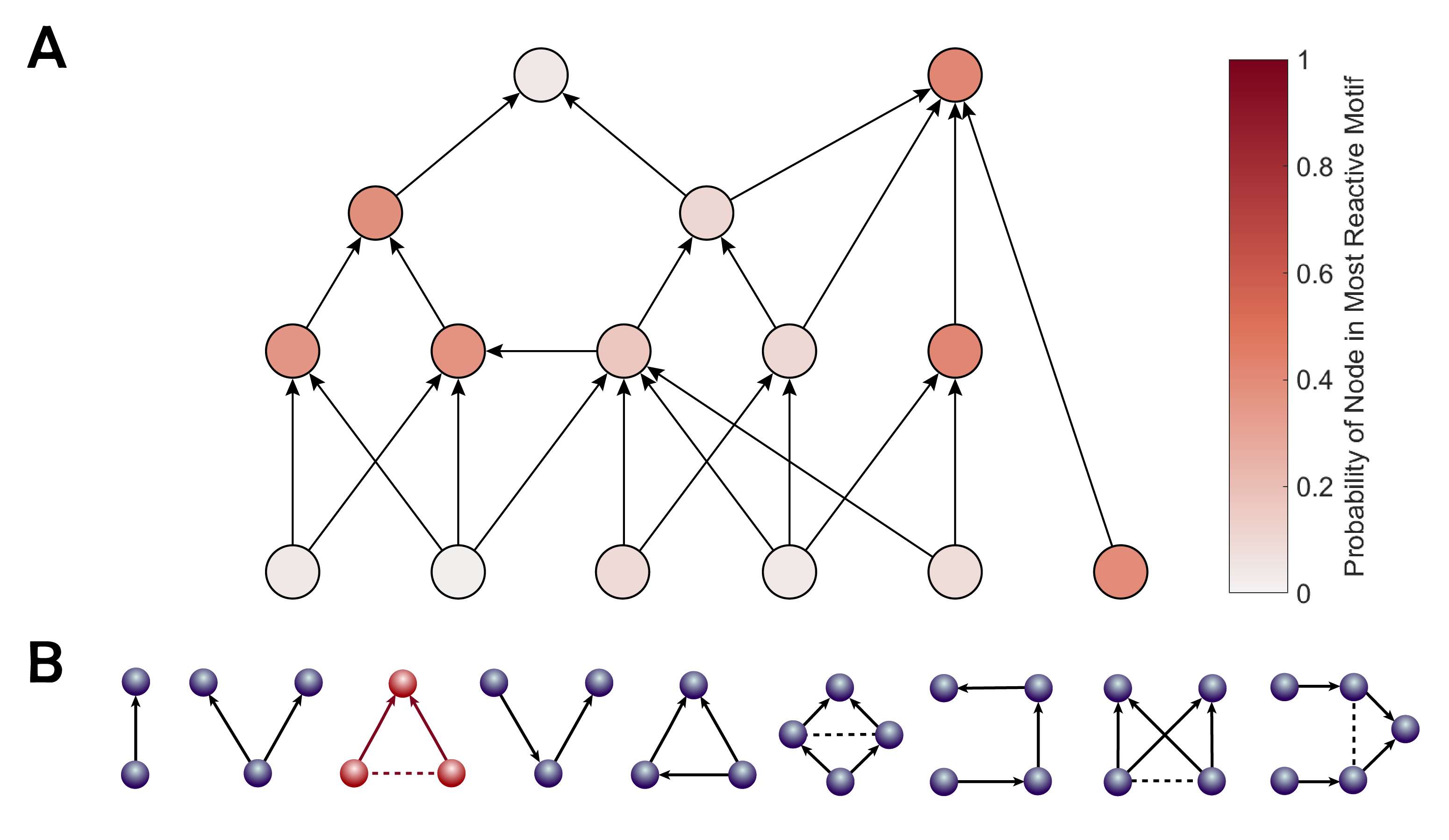}
  \caption{\textbf{Node participation in the most reactive apparent competition motif.} Shown is the example foodweb topology (A).  An overview of other motifs considered in this study is provided below with the apparent competition highlighted (B). In (A), the color of the nodes indicates the probability that a node participates in the most reactive apparent competition motif. Darker shades of red correspond to higher probabilities. The figure shows that the most reactive apparent competition motif most likely consists of the three nodes on the right or the three nodes on the left of the figure. This illustrates that specific network structures drive network reactivity.
}
  \label{fig3:heatmap}
\end{figure}

Since the example topology differed in several ways from the random networks, we examined how its specific structure influences reactivity in more detail (Fig.~4). We focused on the AC motif, as it is the dominant contributor to reactivity in this topology.

In the example network, every node participates in at least one apparent competition motif. However, in $73.43 \%$ of random parameter realizations, the most reactive motif consistently involved one of the same two three-node subgraphs. This suggests that network topology alone can provide valuable insights into the origins of reactivity in a network. 

From these two, the AC motif located in a relatively sparsely connected region on the right side was driving reactivity in this network in $39.05 \%$. Notably, the two links span multiple trophic levels and it features a consumer node that is not regulated by predators. While we find the second reactivity hotspot in a region that at first glance seems better connected, we can see some similarities here when it comes to the regulation of the species involved. Both producers are prey to only one species and the consumer species is as well only regulated by one other species. While we find comparable substructures elsewhere in the network, it is the only instance where both producer species are on intermediate trophic levels. These structural features likely contribute to the motifs' strong reactivities. Exploring whether these characteristics form part of a broader set of rules governing reactivity in networks and foodwebs is a promising target for future research.   

\section*{Conclusions}
Here we examined the conditions under which small parts of a larger network function as predictive motifs. In other words, we studied when analyzing a small part of a network allows for definitive conclusions about the entire system. By integrating insights from theoretical ecology, network science and matrix theory, we arrived at comprehensive answers to this question.

Many important properties of networks, such as dynamical stability, depend on the eigenvalues of certain matrices. In general, considering only a small part of the network does not allow conclusions on the eigenvalues, making it difficult to infer properties of the entire system. However, two notable exceptions exist. First, a motif can determine one or more eigenvalues of the entire system if it exactly or approximately localizes an eigenvector of the full system. This is a relatively rare case, but it explains aspects of the competitive-exclusion principle that have been a driver of ecological insights \cite{gause2019struggle, hardin1960competitive, macarthur1958population}.

Second, and perhaps more importantly, when the relevant matrices are symmetric, every motif is a functional motif. In this case, analyzing any subset of network nodes provides a lower bound for the largest eigenvalue of the entire system. We showed that this second result applies to the study of reactivity of ecological foodwebs. Our numerical experiments further revealed that reactivity of a network is often rooted in small motifs within the larger network.

This result opens up two new avenues of research in foodwebs. It shows for example that it is promising to try to identify topological properties of networks that lead to highly reactive motifs (\emph{e.g.}, Fig.~\ref{fig3:heatmap}). Additionally, our findings imply that it would be valuable to determine the reactivity of key foodweb motifs in field observations and laboratory experiments. Reactivity can be estimated quite easily from time series, but collecting time series from all species in a foodweb is well out of reach for most systems. But, our results show that every subgraph of the web functions as a reactivity motif, and so measuring the reactivity of individual species or specific links between species can already indicate how much risk for violent responses to perturbations originates from that specific part of the network.    

While we have used foodwebs as a primary example, we believe that the results apply to a wide class of dynamical systems that are at risk from perturbations, ranging from power grids and supply chains to networks of social influence. Due to the mathematical properties of reactivity presented here, we can say that in these networks, every subgraph will be a reactivity motif. However, determining how much of a system's reactivity is rooted in small subgraphs in these applications will require subsequent studies. 

We hope future research, perhaps inspired by this work, will deepen our understanding of which system properties arise from parts of a network and which emerge from the network as a whole. In the former case, functional motifs can be identified to narrow the focus to the most critical parts. In the latter case, details should not matter, enabling the use of coarse-grained models. In both cases a simplification is achieved. Further insights into when to pursue one route or the other might ultimately lead to a general theory for the mathematical modeling of complex systems.

\section*{Acknowledgements}
We thank Helge Aufderheide for groundbreaking work on eigenvector localization in the exploitative-competition motif, which he completed as part of his PhD thesis.

\begingroup
\setlength\bibitemsep{0pt}
\printbibliography
\endgroup


\begin{center}
\section*{Supplementary Materials for\\ \scititle}

Melanie~Habermann$^{\ast}$\\ 
\small$^\ast$Corresponding author. Email: melanie.habermann@hifmb.de\\
\end{center}

\subsubsection*{This PDF file includes:}
Materials and Methods\\
Figures S1 to S2\\

\subsubsection*{Other Supplementary Materials for this manuscript:}
Code Availability
\newpage


\sloppy
\baselineskip24pt
\thispagestyle{fancy}


\noindent

\section*{Eigenvectors and Eigenvector Localization}
A vector $\vec{x}$ can be imagined as a line segment with a specific length and direction. Multiplying $\vec{x}$ with a matrix $\mat{A}$ typically changes both properties and will result in a new vector. The exception to this behavior is a group of vectors called eigenvectors. 

Eigenvectors are all non-zero vectors that retain their direction and only change in length by a factor known as the eigenvalue $\lambda$.
The relation between the matrix $\mat{A}$, the eigenvector $\vec{x}$, and the eigenvalue $\lambda$ is represented by the eigenequation: 
\al{
    \mat{A} \vec{x} = \lambda \vec{x} \ .
}

In a system with $N$ variables, each eigenvector has $N$ elements that define the state of each variable. In general, these eigenstates extend throughout the entire system,  meaning that every individual part of the network can influence its state and with that also its stability. 

But again, there is an exception to this rule. Some eigenvectors can be linked to specific subgraphs and free of any outside influences. We call these eigenvectors localized. Localized eigenvectors react only to changes inside their subgraph and are unaffected by any alterations outside of it.  

For better differentiation, we can divide the network into three different sets of nodes. We define one set $I$ as the subgraph's inner nodes. These nodes are only connected to nodes that are part of the subgraph themselves. The second set $C$ contains the connecting nodes. Connecting nodes belong to the subgraph and have also links to nodes outside of it. And lastly, the set $R$ includes all nodes that are not part of $I$ or $C$ \cite{do2012mesostruct}.  

Localized eigenvectors are associated with symmetric subgraphs and satisfy two conditions: All elements in the groups $R$ and $C$ are zero and the sum of the elements of group $I$ is zero.

The symmetry is part of the network topology that is often described by an adjacency matrix $\mat{A}$. The matrix $\mat{A}$ is a binary representation of links between different nodes or, in the case of foodwebs, between species. For foodwebs $\mat{A}$ is the representation of a directed network. The matrix entries indicate the feeding relationships and are one, if a species $j$ is prey to a species $i$, and zero otherwise:
\al{
    \mat{A} : A_{ij} = \begin{cases}
     1 & \text{if $i$ preys on $j$} \\
     0 & \text{otherwise}
   \end{cases} \ .
}

However, when working with foodwebs we can also use the Jacobian matrix $\mat{J}$ instead of $\mat{A}$. The Jacobian will serve as a weighted version of the adjacency matrix. It is important to note, though, that the two matrices are not an identical description of the interactions inside the foodweb. While the adjacency matrix only considers predator-prey relations between species, we have additional entries in the Jacobian that are a result of indirect mutualism between species that are prey to the same predator. This becomes especially important when working with the apparent competition motif where the effect of changes in one prey's biomass on the other plays a crucial role in the dynamics. 

The condition for a localized eigenvector of the Jacobian matrix can be summarized as:
\al{
    \sum_i J_{ji} v_i = c v_j \begin{cases}
        = 0 & \text{for all $j \in R, C$} \\
        \in \mathbb{C} & \text{for all $j \in I$ }
    \end{cases}
}

\section*{Stability}

Dynamical systems can be described by a set of ordinary differential equations
\al{
    \dv{x}{t} = F(x) \ ,
}
where $\vec{x}$ is a vector containing all state variables and $F(\vec{x})$ are the functions that define the changes.

These kinds of systems reach a steady state $x^*$, when
\al{
    F(x^*) = 0 \ .
}

Ecological systems are usually non-linear in nature. To analyse the stability of such a system we rely on the approximation we get by linearizing near a steady state represented by the Jacobian matrix $\mat{J}$. In a system with $N$ different state variables $\mat{J}$ is a $N \times N$ matrix with
\al{
    \mat{J} : J_{ij} = \pdv{\dot{x}_i}{x_j}
}
where $i, j = 1, \dots, N$.

A steady state is considered asymptotically stable when any sufficiently small perturbations away from it decrease over time. This behaviour is predominantly determined by the Jacobian's largest eigenvalue. 

The general solution for the system is 
\al{
    \vec{x}(t) = c_1 e^{\Lambda_1 t} \vec{v}_1 + c_2 e^{\Lambda_2 t} \vec{v}_2 + \dots + c_n e^{\Lambda_n t} \vec{v}_n
    \label{Eq:solution}
}
where the $\Lambda_i$ are the eigenvalues of the Jacobian matrix and $\vec{v}_i$ the corresponding eigenvectors. The solution is dominated by the term with the largest $\Lambda_i$ or more specifically by its real part as it constitutes the largest part of the sum in Eq.~\ref{Eq:solution}.

If the real part of the largest eigenvalue $Re(\Lambda_{max})$ is smaller than $0$, it means any deviation away from the steady state will damp out over time, and the system will return to it. Otherwise, if $Re(\Lambda_1) > 0$, the disturbance is amplified, and the system diverges further away, making the state unstable. Disturbances in unstable states grow either infinitely or until the system reaches another, stable state. Parameter changes cause a shift in the eigenvalue spectrum and can result in at least one of them getting a positive real part, causing the system to become unstable.

\section*{Reactivity}
Stability can be understood as the system's long-term response to any small perturbation. In this context, reactivity can be seen as the short-term or immediate answer. Reactivity, in essence, measures how a system responds to perturbations by calculating the maximum growth rate immediately following such an event \cite{neubert1997alternatives}. We can define reactivity $r$ as

\al{
    r = \max \frac{1}{\mid \vec{\delta} \mid} \dv{\mid \vec{\delta} \mid}{t}
}
where 
\al{
    \mid \vec{\delta} \mid = \sqrt{\vec{\delta}^T \vec{\delta}} 
}
is the magnitude of the deviation $\vec{\delta}$ away from the steady state.

If the growth rate is negative, the perturbation decays right after it occurred and the system returns to its steady state. Steady states where this is the case are considered to be non-reactive. 

With a positive growth rate the perturbation gets amplified causing the system to move further away from its equilibrium. These states are called reactive. Reactive states can either be unstable or stable. 

Neubert and Caswell~\cite{neubert1997alternatives} showed that $r$ is the same as the largest or leading eigenvalue of the symmetric part $\mat{S}$ of the Jacobian matrix 
\al{
    \mat{S} = \frac{\mat{J} + \mat{J}^T}{2} \ .
    \label{Eq:Hermitian}
}

The creates a real symmetric matrix which is a special case of a Hermitian matrix. All properties and theorems that apply to Hermitian matrices apply here as well. One that we can make use of here is Cauchy’s Interlacing Theorem. Simply put the theorem states that the eigenvalues of a principal submatrix consistently fall between two consecutive eigenvalues of the original Hermitian matrix and alternate in magnitude. Let $\mat{S}$ be a $N \times N$ Hermitian matrix with eigenvalues $\lambda_i$, $i = 1, \ldots n$, and $\mat{\hat{S}}$ its $(N-1) \times (N-1)$ principle minor with eigenvalues $\hat{\lambda}_k$, $k = 1, \ldots, N-1$. We get the following sequence of eigenvalues of both matrices in ascending order:

\al{
    \lambda_n \leq \hat{\lambda}_{n-1} \leq \lambda_{n-1} \leq \ldots \leq \hat{\lambda}_{2} \leq \lambda_{2} \leq \hat{\lambda}_{1}  \leq \lambda_1 \ .
}

This also means that the largest eigenvalue of a submatrix is always positioned between the largest and second-largest eigenvalues of the original matrix. Or simply: The largest eigenvalue of the Hermitian matrix is always the largest compared to the eigenvalues of its principal minors.

This has an interesting implication for our work. To analyze the reactivity of network motifs, we use the principal minors of the Hermitian matrix where we delete all columns and rows corresponding to nodes that are not part of the motif. In terms of reactivity, this implies that transitioning from a small subgraph to the entire network inevitably leads to increased reactivity.

Working with the principle minors also allows for an easy way to determine the number of negative eigenvalues. Since the Hermitian matrix is symmetric by design, we can use the Sylvester criterion. The criterion states that the number of negative eigenvalues is the same as the number of sign changes in the sequence of principle minors. 

To create this sequence we remove one row and the corresponding column that belong to a species that is not part of the smaller motif in each step and calculate the matrix's determinant. As an example, we can look at the sequence of principle minors for a $3 \times 3$ symmetric matrix $\mat{S}$:
\al{
     \left\vert 
     \begin{array}{c c c} S_{11} & S_{12} & S_{13} \\ 
     S_{12} & S_{22} & S_{23} \\ 
     S_{13} & S_{23} & S_{33}
    \end{array} 
    \right\vert, 
    \left\vert 
    \begin{array}{c c} S_{11} & S_{12} \\ 
    S_{12} & S_{22} \\
    \end{array}
    \right\vert, 
    \left\vert S_{11} \right\vert, 1 
}
where the last element (the determinant of size 0) is 1 by definition. 

Starting from small one node motifs, we are able to identify conditions where a motif becomes reactive (given the smaller motifs are considered to be non-reactive). A one node motif becomes reactive, if 
\al{
    \left\vert S_{11} \right\vert > 0 \ .
}

This has to be the case because a non-reactive motif would require one negative eigenvalue, so we would need a sign change between the Stephan from $1$ to $\left\vert S_{11} \right\vert$. If we increase the size of the motif by another species and assume that both species are non-reactive in isolation, the determinant of the $2 \times 2$ motif would need to be negative for the system to be reactive. Following this logic we can identify the conditions needed for a motif to switch between a reactive and non-reactive state.

\section*{Generalized Foodweb Model}

This section will only superficially explain the concept of Generalized Modeling based on the model used in this study. A more detailed explanation of Generalized Modeling and further use cases in different scientific disciplines can be found in \cite{massing2021gm}.

The processes involved in the dynamics will not be restricted to specific functional forms but are kept as general as possible. This way we are able to represent a whole set of different models. To achieve this we use a Generalized Modeling approach for foodweb models based on the model in \cite{gross2006gm}. Consider a system with $N$ different populations of species $X_1$, $X_2$ \dots, $X_N$. We can describe the dynamics of each population by an ODE of the form
\al{
    \dot{X}_i = S_i(X_i) + \eta_i F_i(\vec{X}) - M_i(X_i) - \sum_{j = 1}^N L_{ji}(\vec{X}) \ ,
    \label{Eq:GMFoodWeb}
}
where $X_i$ represents the abundance of the species in question and $S_i(X_i)$, $F_i(\vec{X})$, $M_i(X_i)$, and $L_{ji}(X_i)$ are non-linear functions for the population gains and losses. Depending on the species, population growth is either due to primary production $S(X)$ or predation on other species $F(X)$. Since the consumed food is not entirely converted to a population gain for the predator, we include a conversion factor $\eta$. Population losses can either be caused by natural mortality (e.g., disease, age) or predation by a different species $j$. We do not consider cannibalistic behaviour in this study.

A predator-prey relationship necessitates a link between the two functions $F_i(\vec{X})$ and $L_{ji}(\vec{X})$. Additionally, it should account for the option that a predator can consume multiple prey species. To represent this more realistically, we introduce two auxiliary variables, $T_j(\vec{X})$ and $C_{ji}(X_j)$. $T(X)$ determines the overall food supply for species $j$ whereas $C_{ji}(X_j)$ defines the contribution of species $i$ in the diet of $j$. We can define
\al{
    T_j(\vec{X}) = \sum_{i = 1}^N C_{ji}(X_i)\ .
}

By employing $T_j(\vec{X})$ we can redefine $F_j(\vec{X})$ as $F_j(T_j(\vec{X}), X_j)$.

Now, when we factor in the portion of the total food supply that species $i$ contributes to the predator's diet, we can modify the predation loss function as follows:
\al{
    L_{ji}(\vec{X}) = \frac{C_{ji}(X_j)}{T_j(\vec{X})} F_j(T_j(\vec{X), X_j)}\ .
}

\subsection*{Normalizing the Model}

For a model with the complexity of a foodweb we can be sure of at least one steady state $\vec{X}*$, where each state variable has its equilibrium value $X_i^*$. To simplify things, we use this to define normalized state and auxiliary variables as follows
\al{
    x_i &= \frac{X_i}{X_i^*}\ , \\
    t_j(\vec{x}) &= \frac{T_j(\vec{X})}{T_j^*}\ , \\
    c_{ji}(x_i) &= \frac{C_{ji}(X_i)}{C_{ji}^*}\ .
}

We also introduce normalized versions of each process function $P_i(X_i)$ involved in the model:
\al{
    p_i(x_i) = \frac{P_i(X_i)}{P_i(X_i^*)}\ ,
}
where $P_i(X_i^*)$ is the process function in the steady state. For better readability in the following parts we use $P_i^*$ instead of $ P_i(X_i^*)$.

Rewriting the model from Eq.~\ref{Eq:GMFoodWeb} with these normalized quantities yields
\al{
    \dot{x}_i = \frac{S_i^*}{X_i^*} s_i(x_i) + \frac{F_i^*}{X_i^*} f_i(\vec{x}, x_i) - \frac{M_i^*}{X_i^*} m_i(x_i) - \sum_{j = 1}^N \frac{L_{ji}^*}{X_i^*} l_{ji}(\vec{x})\ ,
    \label{Eq:normmod}
}
where 
\al{
    l_{ji}(\vec{x}) = \frac{C_{ji}^* F_j^*}{T_j^* L_{ji}^*} \frac{c_{ji}}{t_j} f_j(t_j(\vec{x}, x_j) = \frac{c_{ji}}{t_j} f_j(t_j(\vec{x}, x_j)\  .
}

For the total food supply $t_j(\vec{x})$, we get
\al{
    t_j(\vec{x}) = \sum_{i = 1}^N \frac{C{ji}^*}{T_j^*} c_{ji}(x_i)\ .
    \label{Eq:normtotalamount}
}

\subsection*{Scale Parameters}

The model can be further simplified by introducing easily interpretable scale parameters. Scale parameters quantify the biomass flow either of a whole population or define the proportions of the processes involved in the gains and losses. 

If we consider the system in its steady state, $\dot{\vec{x}}$ is by definition zero. Because we normalized the model in the previous steps, all state variables and process rates are now one. If we take this into consideration, we can write the model in its steady state as
\al{
    0 = \frac{S_i^*}{X_i^*} + \frac{F_i^*}{X_i^*} - \frac{M_i^*}{X_i^*} - \sum_{j = 1}^N \frac{L_{ji}^*}{X_i^*}\ .
    \label{Eq:steadystate}
}
We are left with a sum of constant values that we use as our scale parameters. 

For the overall biomass flow of species $i$, we define a parameter
\al{
    \alpha_{xi} = \frac{S_i^*}{X_i^*} + \frac{F_i^*}{X_i^*} - \frac{M_i^*}{X_i^*} - \sum_{j = 1}^N \frac{L_{ij}^*}{X_i^*}\ .
}
This will leave us with a number of $N$ parameters just for the overall biomass flow of all $N$ species. We can reduce this number by one, if we rescale the time for all species with the turnover rate of the species on the lowest trophic level $\alpha_1$.

The two gain terms for primary production and predation account for the positive biomass flow. To quantify the relative contributions of each process we use 
\al{
    \rho_{xi} &= \frac{1}{\alpha_{xi}} \frac{S_i^*}{X_i^*}\ , \\
    \hat{\rho}_{xi} &= 1 - \rho_{xi} = \frac{1}{\alpha_{xi}} \frac{F_i^*}{X_i^*}\ .
}
We also define parameters for the relative contribution to the biomass loss
\al{
    \sigma_{xi} &= \frac{1}{\alpha_{xi}} \frac{M_i^*}{X_i^*}\ , \\
    \hat{\sigma}_{xi} &= \frac{1}{\alpha_{xi}} \sum_{j = 1}^N \frac{L_{ij}^*}{X_i^*}
}

For species that are hunted by multiple predators the proportional loss caused by each predator is represented by additional branching parameters 
\al{
    \beta_{ji} = \frac{1}{\alpha_{xi} \hat{\sigma}_{xi}} \frac{L_{ij}^*}{X_i^*}\ .
}

In Eq.~\ref{Eq:normtotalamount} we already saw that each prey species contributions only a fraction 
\al{
    \frac{C_{ji}^*}{T_j^*}\ 
}
to a predators diet. Since these are also constant values, we redefine them as additional scale parameters
\al{
    \chi_{ji} = \frac{C_{ji}^*}{T_j^*}\ ,
}
With this Eq.~\ref{Eq:normtotalamount} becomes
\al{
    t_j(\vec{x}) = \sum_{i = 1}^N \chi_{ji} c_{ji}(x_i)\ .
}

Including the set of scale parameters in Eq.~\ref{Eq:steadystate} leaves us with the functions
\al{
    \dot{x}_i = \alpha_i \left( \rho_{xi} s_i(x_i) + \hat{\rho}_{xi} f_i(\vec{x}, x_i) - \sigma_{xi} m_i(x_i) - \hat{\sigma}_{xi} \sum_{j = 1}^N \beta_{ji} l_{ji}(\vec{x}) \right)\ .
}
where
\al{
    l_{ji}(\vec{x}) = \frac{c_{ji}}{t_j} f_j(t_j(\vec{x}, x_j)\ , \\
    t_j(\vec{x}) = \sum_{i = 1}^N \chi_{ji} c_{ji}(x_i)\ .
}

To reduce the number of parameters in the following calculations we assume that predation is the main cause of death in prey species and affects the species to a significantly larger extent, so that $m_i(x_i)$ can be neglected in these cases. For this, we set $\sigma_{xi} = 0$ and $\hat{\sigma}_{xi} = 1$ for all prey populations. Only for top predators, we assume the opposite to be true.

We can also savely assume that primary production is a trait we will only find on the lowest trophic level. To represent this in our model we set $\rho_{xi} = 1$ and $\hat{\rho}_{xi} = 0$ for primary producers and vice versa for every species on a higher trophic level. 

\subsection*{Computation of the Jacobian and Corresponding Symmetric Matrix}

To compute the Hermitian matrix we first have to construct the system's Jacobian. We introduce another set of parameters (i.e., exponent parameters) that will represent the partial derivatives of out general process functions. Again, you can find sensible interpretations for these parameters. In general, we can understand them as the non-linearity of the corresponding processes. 

The new set of parameters include
\al{    
    \nonumber s_{xi} &:= \left. \pdv{s(x_i)}{x_i} \right\vert_* , & 
    f_{xi} &:= \left. \pdv{f(t_i(\vec{x}), x_i)}{x_i} \right\vert_* , &
    f_{tj} &:= \left. \pdv{f(t_j(\vec{x}), x_j)}{t_j} \right\vert_* , \\
    \lambda_{ji} &:= \left. \pdv{c_{ji}}{x_i} \right\vert_* , &
    \text{and} \  m_{xi} &:= \left. \pdv{m(x_i)}{x_i} \right\vert_*\ .
    \label{Eq:ExpoPara}
}

Now that we defined the exponent parameters we can use them to construct our Jacobian matrix $\mat{J}$. The resulting Jacobian has the diagonal entries
\al{
    \mat{J}_{ii} = \alpha_i \left( \rho_{xi} s_{xi} + \hat{\rho}_{xi} f_{xi}  - \sigma_{xi} m_{xi} - \hat{\sigma}_{xi} \sum_{j = 1}^N \beta_{ji} \lambda_{ji} (\chi_{ji}(f_{tj} - 1) + 1) \right)
}
and non-diagonals
\al{
    \mat{J}_{ni} = \alpha_n \left( \hat{\rho}_{xn}\lambda_{ni}\chi_{ni}f_{tn} - \hat{\sigma}_{xn} \sum_{j = 1}^N \beta_{jn} \lambda_{ji} \chi_{ji}(f_{tj} - 1) \right) \ .
}

Following the step outlined in Eq.~\ref{Eq:Hermitian} we can calculate the Hermitian matrix. It is important to note that the diagonal entries of $\mat{S}$ are identical to those in the Jacobian. These diagonal elements indicate the effect that a species has on itself, and they provide insights into whether the network contains reactive nodes or not. If we can already identify reactive nodes (i.e., positive values on the main diagonal), we can safely say that the system is reactive as well. This is important information in itself, although node reactivity didn't prove to be a good approximation for total system reactivity. It needs at least the interaction between two nodes to provide a good measure of a lower bound.

\clearpage

\section*{Histograms of additional motifs}

\begin{figure}[ht!]
  \centering
  \includegraphics[width=0.9\textwidth]{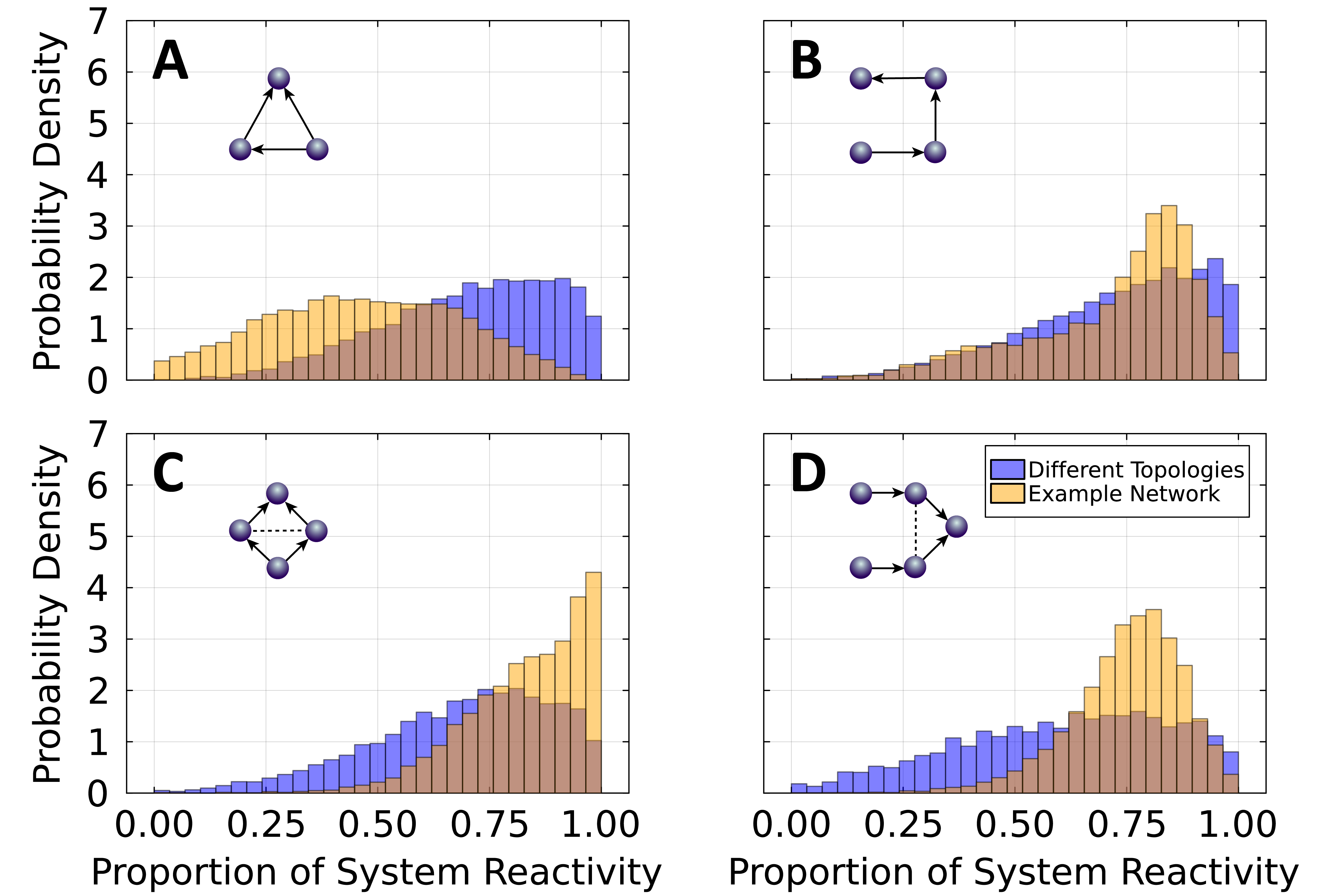}
  \caption{\textbf{Contribution of individual motifs to system reactivity for larger motifs.} The histograms compare the proportion of total system reactivity explained by the most reactive instance of a single motif in $10^4$ different communities. The motifs in question are the three-node omnivory motif (O, panel A), the tetra-trophic foodchain (FC, B), the diamond (D, C), and the combined tri-trophic foodchain (CTC). The structure of each motif is presented in the upper left corner. For the 15-node example network with a fixed topology (orange), the D motif accounts for the highest proportions of total system reactivity, reaching $\geq 85 \%$ with a higher probability than any of the other shown motifs. In the set of networks with differing topologies (blue), O, FC, and D provide similar results with relatively strong contributions to system reactivity. The TCT motif was better suited to explain reactivity in the example network than it was in the case of the ensemble of networks, highlighting the importance of analysing network structure before deciding on a motif as a predictor for total system reactivity.}
  \label{figS1:comp_motifs}
\end{figure}

\subsection*{Code Availability}

The code used to perform the analyses is available at https://github.com/MelHabm/FunctionalMotifsInFoodwebs. 

\end{document}